\documentclass[letterpaper,twocolumn,fleqn]{article} 

\usepackage{ist}
\usepackage{algorithm}
\usepackage{algpseudocode}
\usepackage{amsmath}
\usepackage{amssymb}
\usepackage{mathtools}

\algnewcommand\algorithmicinput{\textbf{Input:}}
\algnewcommand\Input{\item[\algorithmicinput]}

\DeclareMathOperator*{\argmin}{arg\!\min}

\title{Single Shot Digital Holography Using Iterative Reconstruction with Alternating Updates of Amplitude and Phase}

\author{Dennis J. Lee$^{1,2,\ast}$, Charles A. Bouman$^2$, and Andrew M. Weiner$^2$\\
$^1$Sandia National Laboratories, 1515 Eubank Blvd. SE, Albuquerque, NM 87123, USA\\
$^2$School of Electrical and Computer Engineering, Purdue University, 465 Northwestern Avenue,
West Lafayette, IN 47907, USA\\
$^\ast$Corresponding author: dlee1@sandia.gov}

\date{} 

\begin{document} 

\maketitle 

\thispagestyle{empty} 


\begin{abstract}
We present an image recovery approach to improve amplitude and phase reconstruction from single shot digital holograms, using iterative reconstruction with alternating updates. This approach allows the flexibility to apply different priors to amplitude and phase, improves phase reconstruction in image areas with low amplitudes, and does not require phase unwrapping for regularization. Phantom simulations and experimental measurements of a grating sample both demonstrate that the proposed method helps to reduce noise and resolve finer features. The improved image reconstruction from this technique will benefit the many applications of digital holography.
\end{abstract}

\section{Introduction}
Digital holography has many versatile applications, including microscopy \cite{Yamaguchi01}, phase contrast \cite{Xu01}, 3D displays \cite{Matoba02}, tomography \cite{Choi07}, and terahertz imaging \cite{Heimbeck11}.  This technique enables the measurement of both amplitude and phase, useful for quantifying path lengths, measuring index contrast, or viewing biological samples \cite{Lee14, Lee15}. Various experimental methods can extract amplitude and phase from a measured interference pattern: for example, phase-shifting interferometry involves incrementally stepping the phase of the reference beam, from which the object phase can be deduced \cite{Yamaguchi97}. However, this method requires multiple images to be recorded on a vibration-free optical table, typically with expensive devices such as high frame rate cameras.

Off-axis interferometry, based on the interference of an object and reference beam slightly offset in angle, can compute amplitude and phase with a single measurement. Conventional processing spatially filters the Fourier transform of the hologram \cite{Weiner09}. While simple, Fourier filtering suffers from some drawbacks: The filter window size is subjective, and
good reconstruction requires the zero order and cross terms to be well-separated. A large separation between object and reference beams results in a high carrier frequency hologram, in which the zero-order and sidebands are well-separated in the spectrum. Although a high carrier frequency is ideal for Fourier filtering, other factors may prevent perfect reconstruction: The experimental configuration may not allow a large angular separation of the object and reference beams, the sampling requirements of the camera pixels may limit the angular separation, or the object may contain high spatial frequencies.

To overcome these drawbacks, other approaches include subtracting the zero order term from the hologram \cite{Chen07}, iteratively solving for the field in the spatial domain \cite{Ma09} and the frequency domain \cite{Pavillon10}, and applying a nonlinear filter \cite{Pavillon09}.  Optimization-based approaches include formulating holography as a nonlinear least squares problem \cite{Liebling04}, as constrained optimization \cite{Khare13}, as penalized likelihood with simulated data \cite{Sotthivirat04}, or as a nonlinear inverse problem \cite{Bourquard13}.

In this work, we propose a new image recovery approach that calculates an object field from a single hologram, using iterative reconstruction with alternating updates of amplitude and phase. To our knowledge, this work is the first application of the alternating update strategy to single shot digital holography. This approach offers multiple advantages:
\begin{enumerate}
\item It allows prior knowledge, such as object smoothness, to be applied separately to amplitude and phase. For example, although nearly transparent biological samples like cells are smooth in amplitude, they exhibit sharp edges in phase.

\item Regularizing phase separately, rather than regularizing the field $Ae^{i\phi}$, mitigates the effects of poor signal areas caused by low amplitudes, aiding phase reconstruction.

\item Our algorithm regularizes phase without requiring phase unwrapping, as discussed in the Theory section.
\end{enumerate}

This work is organized as follows.  In the Theory section, we present continuous and discrete mathematical models of holographic measurements, followed by an iterative reconstruction algorithm that recovers amplitude and phase via alternating updates. In the Experiment section, we discuss results from a phantom simulation and from experimentally measured holograms of a grating sample. Finally we provide some concluding remarks.

\section{Theory}
\subsection{Continuous Model}
In digital holography, an object field, $ o(\mathbf{x}) = A(\mathbf{x}) e^{i \phi(\mathbf{x})} $, and a reference field, $ r(\mathbf{x}) = \tilde{A}(\mathbf{x}) e^{i \tilde{\phi}(\mathbf{x})} $, combine to form an interference pattern measured on a camera,
%
%
%
\begin{equation}
\begin{split}
I_{\text{ideal}}(\mathbf{x}) &= | o(\mathbf{x}) + r(\mathbf{x}) |^2 \\
&= \left| A(\mathbf{x}) e^{i \phi(\mathbf{x})} + \tilde{A}(\mathbf{x}) e^{i \tilde{\phi}(\mathbf{x})} \right|^2 \\ 
&=  A^2(\mathbf{x}) + \tilde{A}^2(\mathbf{x}) + 2 A(\mathbf{x}) \tilde{A}(\mathbf{x}) \text{cos}\left[ \phi(\mathbf{x}) - \tilde{\phi}(\mathbf{x}) \right],
\end{split}
\label{eqn:e_intensity}
\end{equation}
where $ \mathbf{x} \in \mathbb R^2 $ is the spatial coordinate in the camera plane, and $ \tilde{\phi}(\mathbf{x}) $ typically represents linear phase.  The measured hologram is generally noisy, which we can model as Gaussian noise that corrupts the interference pattern $ I_{\text{ideal}}(\mathbf{x}) $.  Let $ I(\mathbf{x}) $ denote the measured hologram.  It is also possible to model the measured hologram using a Poisson distribution \cite{Sauer93, Bouman96}.  We formulate the problem as the minimization of a cost function with the general form
$$
c(A, \phi) = L(A, \phi) + \beta_A R(A) + \beta_\phi R(\phi),
$$ 
where $ L(A, \phi) $ is the negative log-likelihood function corresponding to our model of the noisy hologram as a Gaussian distribution, $ \beta_A $ and $ \beta_\phi $ are scalar regularization parameters, and $ R(A) $ and $ R(\phi) $ are roughness penalty functions for amplitude and phase, respectively.  In the continuous formulation, the likelihood function has the form
\begin{equation}
\begin{split}
L(A, \phi) &= \frac{1}{2} \int_{\mathbb{R}^2} \left\{ A^2(\mathbf{x}) + \tilde{A}^2(\mathbf{x}) \right.\\
&~~~~\left.+ 2 A(\mathbf{x}) \tilde{A}(\mathbf{x}) \text{cos}\left[ \phi(\mathbf{x}) - \tilde{\phi}(\mathbf{x}) \right] - I(\mathbf{x}) \right\}^2 d\mathbf{x}.
\end{split}
\end{equation}
To reduce noise and impose prior knowledge such as edge-preserving smoothness on the object, we also minimize the total variation of the amplitude and phase with the regularizer terms
\begin{equation}
R(A) = \int_{\mathbb{R}^2} \sqrt{| \nabla A(\mathbf{x}) |^2 + \epsilon} ~d\mathbf{x}
\end{equation}
and
\begin{equation}
R(\phi) = \int_{\mathbb{R}^2} \sqrt{| \nabla e^{i \phi(\mathbf{x})} |^2 + \epsilon} ~d\mathbf{x},
\end{equation}
where $ \epsilon $ is a small constant to ensure differentiability.  Other prior models can also be used \cite{Bouman93}.  Now the cost function becomes 
\begin{equation}
\begin{split}
c(A, \phi) &= L(A, \phi) + \beta_A R(A) + \beta_{\phi} R(\phi) \\
&= \frac{1}{2} \int_{\mathbb{R}^2} \left\{ A^2(\mathbf{x}) + \tilde{A}^2(\mathbf{x}) \right.\\
&~~~~\left.+ 2 A(\mathbf{x}) \tilde{A}(\mathbf{x}) \text{cos}\left[ \phi(\mathbf{x}) - \tilde{\phi}(\mathbf{x}) \right] - I(\mathbf{x}) \right\}^2 d\mathbf{x} \\
&~~~~+ \beta_A \int_{\mathbb{R}^2} \sqrt{| \nabla A(\mathbf{x}) |^2 + \epsilon} ~d\mathbf{x}\\
&~~~~+ \beta_{\phi} \int_{\mathbb{R}^2} \sqrt{| \nabla e^{i \phi(\mathbf{x})} |^2 + \epsilon} ~d\mathbf{x}.
\end{split}
\end{equation}
Our goal is to solve for amplitude and phase by minimizing this cost function:
\begin{equation}
\left(\hat{A}, \hat{\phi}\right) = \argmin_{A, \phi} c(A, \phi).
\end{equation}
%

\subsection{Discrete Model}
We discretize the amplitude $\mathbf{A} \in \mathbb R^N$, phase $ \phi \in \mathbb R^N$, and intensity $\mathbf{I} \in \mathbb R^N$ using vector notation:
\begin{equation}
\mathbf{A} = \left[
\begin{array}{ccccc}
A_1 & \dots & A_i & \dots & A_N
\end{array}
\right]^T,
\end{equation}
\begin{equation}
\mathbf{\phi} = \left[
\begin{array}{ccccc}
\phi_1 & \dots & \phi_i & \dots & \phi_N
\end{array}
\right]^T,
\end{equation}
and
\begin{equation}
\mathbf{I} = \left[
\begin{array}{ccccc}
I_1 & \dots & I_i & \dots & I_N
\end{array}
\right]^T,
\end{equation}
where $N$ is the total number of pixels. We use similar vector notations for the reference amplitude and phase $ \tilde{\mathbf{A}} $ and $ \tilde{\phi} $.
According to the pixel pitch $ \Delta_x $ of the camera, we specify the sampled likelihood function as
\begin{equation}
\begin{split}
L(A, \phi) &= \frac{1}{2} \sum_{\mathbf{m} \in \mathbb{Z}^2} \left\{ A^2(\mathbf{x}) + \tilde{A}^2(\mathbf{x}) \right.\\
&~~~~\left.\left.+ 2 A(\mathbf{x}) \tilde{A}(\mathbf{x}) \text{cos}\left[ \phi(\mathbf{x}) -  \tilde{\phi}(\mathbf{x}) \right] - I(\mathbf{x}) \right\}^2 \right|_{\mathbf{x} = \mathbf{m} \Delta_x} \\
&= \frac{1}{2} \sum_{i} \left[ A^2_i + \tilde{A}^2_i + 2 A_i \tilde{A}_i \text{cos}\left( \phi_i - \tilde{\phi}_i \right) - I_i \right]^2.
\end{split}
\end{equation}
We regularize the unknowns by minimizing the total variation, with
\begin{equation}
R(\mathbf{A}) = \sum_{i} \sqrt{\left(\mathbf{C} \mathbf{A} \right)_i^2 + \epsilon}
\end{equation}
and
\begin{equation}
R(\phi) = \sum_{i} \sqrt{\left| \left[ \mathbf{C} e^{i \phi} \right]_i \right|^2 + \epsilon}.
\end{equation}
Here we use the shorthand vector notation
\begin{equation}
e^{i \phi} = \left[
\begin{array}{ccccc}
e^{i \phi_1} & \dots & e^{i \phi_i} & \dots & e^{i \phi_N}
\end{array}
\right]^T.
\end{equation}
$ \mathbf{C} $ is a convolution matrix implementing the discretized first derivatives of the nearest neighbors. For example, for
\begin{equation}
\mathbf{C} = \left[  
\begin{array}{c}
\mathbf{C}_h\\
\mathbf{C}_v
\end{array}
\right],
\end{equation}
$ \mathbf{C} \in \mathbb R^{2N \times N} $ is a concatenation of $ \mathbf{C}_h $ and $ \mathbf{C}_v $, which implements first derivatives in the horizontal and vertical directions.  Since we are examining derivatives of $ e^{i \phi} $, no phase unwrapping is required for regularization.  The cost function becomes
\begin{equation}
\begin{split}
c(\mathbf{A}, \phi) &= L(\mathbf{A}, \phi) + \beta_A R(\mathbf{A}) + \beta_{\phi} R(\phi) \\
&= \frac{1}{2} \sum_{i} \left[ A^2_i + \tilde{A}^2_i + 2 A_i \tilde{A}_i \text{cos}\left( \phi_i - \tilde{\phi}_i \right) - I_i \right]^2 \\
&~~~~+ \beta_A \sum_{i} \sqrt{\left(\mathbf{C} \mathbf{A} \right)_i^2 + \epsilon}
+ \beta_\phi \sum_{i} \sqrt{\left| \left[ \mathbf{C} e^{i \phi} \right]_i \right|^2 + \epsilon}.
\end{split}
\label{eqn:e_c_Aphi}
\end{equation}
Our goal is to solve for $ \mathbf{A} $ and $ \mathbf{\phi} $ by minimizing this cost function:
\begin{equation}
\left(\hat{\mathbf{A}}, \hat{\phi}\right) = \argmin_{\mathbf{A}, \phi \in \mathbb{R}^N} c(\mathbf{A}, \phi).
\label{eqn:cost}
\end{equation}
We estimate $ \mathbf{A} $ and $ \phi $ by alternatively updating them in each iteration as
\begin{equation}
\mathbf{A}^{(n+1)} = \argmin_{\mathbf{A} \in \mathbb{R}^N} c(\mathbf{A}, \phi^{(n)})
\label{eqn:argmin_A}
\end{equation}
and
\begin{equation}
\phi^{(n+1)} = \argmin_{\phi \in \mathbb{R}^N} c(\mathbf{A}^{(n+1)}, \phi).
\label{eqn:argmin_phi}
\end{equation}
\begin{algorithm}
\begin{algorithmic}[1]
\Statex \textbf{Input:} Measured hologram
\Statex \textbf{Outputs:} Amplitude $\mathbf A$ and phase $\phi$
\Statex
\State $ n \gets 0 $
\State Initialize $ \mathbf A^{(0)} $ and $ \mathbf \phi^{(0)} $ by Fourier filtering.
\While{$c(\mathbf A^{(n)}, \phi^{(n)}) > \epsilon_1$}
\Statex
\State $ i \gets 0 $
\State $ \mathbf A^{(n, ~0)} \gets \mathbf A^{(n)} $
\While{$c(\mathbf A^{(n, ~i)}, \phi^{(n)}) > \epsilon_2$}
\State $ \left. \mathbf d \gets -\nabla_{\mathbf A} c(\mathbf A, \phi^{(n)})  \right|_{\mathbf{A} = \mathbf{A}^{(n,~i)}} $ \par
\hskip \algorithmicindent $~~= \left. \left. -\nabla_{\mathbf A} L(\mathbf A, \phi^{(n)} )  \right|_{\mathbf{A} = \mathbf{A}^{(n,~i)}}
- \beta_A \nabla_{\mathbf A} R( \mathbf A ) \right|_{\mathbf{A} = \mathbf{A}^{(n,~i)}} $
\State $\alpha^\ast \gets \argmin_{\alpha \in \mathbb R} ~c\left(\mathbf A^{(n, ~i)} + \alpha \mathbf d, \phi^{(n)} \right) $
\State $ \mathbf A^{(n, ~i + 1)} \gets \mathbf A^{(n,~i)} + \alpha^\ast \mathbf d $
\State $ i \gets i + 1 $
\EndWhile
\Statex
\State $\mathbf A^{(n+1)} \gets \mathbf A^{(n, ~i)}$
\Statex
\State $ i \gets 0 $
\State $ \phi^{(n, ~0)} \gets \phi^{(n)} $
\While{$c(\mathbf A^{(n+1)}, \phi^{(n, ~i)}) > \epsilon_3$}
\State $ \left. \mathbf d \gets -\nabla_{\phi} c(\mathbf A^{(n+1)}, \phi)  \right|_{\phi = \phi^{(n,~i)}} $ \par
\hskip \algorithmicindent $~~= \left. \left. -\nabla_\phi L(\mathbf A^{(n + 1)}, \phi )  \right|_{\phi = \phi^{(n,~i)}}
- \beta_\phi \nabla_{\phi} R( \phi ) \right|_{\phi = \phi^{(n,~i)}} $
\State $\alpha^\ast \gets \argmin_{\alpha \in \mathbb R} ~c\left(\mathbf A^{(n + 1)}, \phi^{(n, ~i)} + \alpha \mathbf d \right) $
\State $ \phi^{(n, ~i + 1)} \gets \phi^{(n, ~i)} + \alpha^\ast \mathbf d $
\State $ i \gets i + 1 $
\EndWhile
\Statex
\State $ \phi^{(n + 1)} \gets \phi^{(n, ~i)} $
\Statex
\State $n \gets n + 1$
\EndWhile
\Statex
\State $ \mathbf A \gets \mathbf A^{(n)} $
\State $ \phi \gets \phi^{(n)} $
\State \Return $\mathbf A$, $\phi$
\end{algorithmic}
\caption{Iterative Reconstruction with Alternating Updates of Amplitude and Phase}
\label{IR_Algorithm}
\end{algorithm}
\begin{table}
\centering
\begin{tabular}{|c|c|c|}
\hline
\textbf{Reconstruction} & \textbf{Simulation} & \textbf{Experimental}\\
\textbf{Techniques} & \textbf{Results} & \textbf{Results}\\
\hline
Fourier Filtering & Figure \ref{Figure:Phantom_Reconstruction}(b) & Figure \ref{Figure:Hologram_Reconstruction}(b) \\
\hline
Iterative Reconstruction & Figure \ref{Figure:Phantom_Reconstruction}(c) & Figure \ref{Figure:Hologram_Reconstruction}(c)\\
\hline
\end{tabular}
\begin{center}
\caption{$~~~~~~~$Table 1. Comparison of reconstruction techniques.}
\end{center}
\label{tab:reconstruction}
\end{table}
\begin{table}
\centering
\begin{tabular}{|c|c|}
\hline
\textbf{Experimental Data} & \textbf{Corresponding Figure} \\
\hline
Hologram with a & Figure \ref{Figure:Hologram}(a) \\
low carrier frequency &  \\ 
\hline
Hologram with a & Figure \ref{Figure:Hologram}(d) \\
high carrier frequency &  \\
\hline
\end{tabular}
\begin{center}
\caption{$~~~~~~~$Table 2. Experiment data and corresponding figures.}
\end{center}
\label{tab:experimental_data}
\end{table}
%
%

\subsection{Iterative Reconstruction Algorithm}
To solve the reconstruction problem in Eq. (\ref{eqn:cost}), we employ the alternating update strategy in Eqs. (\ref{eqn:argmin_A}) - (\ref{eqn:argmin_phi}), detailed in Algorithm \ref{IR_Algorithm}. First fixing $\phi$, we update $\mathbf A$ through a gradient descent approach \cite{Boyd04}, until the cost function reduces to a specified threshold $\epsilon_2$; updates for $\phi$ proceed similarly, using the new value of $ \mathbf A $. The alternating updates repeat, enclosed in an outer loop, terminating when cost reduces to a threshold $ \epsilon_1 $.

The update steps require various derivatives to be computed. For reference, we list the gradients of the likelihood functions,
\begin{equation}
\begin{split}
\left[ \nabla_{\mathbf{A}} L(\mathbf{A}, \phi) \right]_i &=  \left[ A_i^2 + \tilde{A}_i^2 + 2 A_i \tilde{A}_i ~\text{cos}\left(\phi_i - \tilde{\phi}_i\right) - I_i \right]\\
&\cdot \left[2 A_i + 2 \tilde{A}_i ~\text{cos}\left(\phi_i - \tilde{\phi}_i \right)\right]
\end{split}
\end{equation}
and
\begin{equation}
\begin{split}
\left[ \nabla_{\phi} L(\mathbf{A}, \phi) \right]_i = &- \left[ A_i^2 + \tilde{A}_i^2 + 2 A_i \tilde{A}_i ~\text{cos}\left(\phi_i - \tilde{\phi}_i\right) - I_i \right]\\
 &\cdot \left[2 A_i \tilde{A}_i ~\text{sin}\left(\phi_i - \tilde{\phi}_i \right)\right],
\end{split}
\end{equation}
and the Laplacians of the likelihood functions, diagonal matrices with entries
\begin{equation}
\begin{split}
\left[ \nabla^2_{\mathbf{A}} L(\mathbf{A}, \phi) \right]_{ii} = &\left[2 A_i + 2 \tilde{A}_i 
\text{cos}(\phi_i - \tilde{\phi}_i) \right]^2 \\
&+ 2  \left[A_i^2 + \tilde{A}_i^2 + 2  A_i  \tilde{A}_i
 \text{cos}(\phi_i - \tilde{\phi}_i) - I_i\right]
\end{split}
\end{equation}
and
\begin{equation}
\begin{split}
\left[ \nabla^2_\phi L(\mathbf A, \phi) \right]_{ii} = &\left[ 2 A_i \tilde{A}_i \text{sin}(\phi -
\tilde{\phi}) \right]^2 \\
&+ \left\{I_i - \left[A_i^2 + \tilde{A}_i^2 + 2 A_i \tilde{A}_i \text{cos}\left(\phi_i -
\tilde{\phi}_i\right)\right]\right\}\\
&\cdot 2 A_i \tilde{A}_i \text{cos}(\phi_i - \tilde{\phi}_i).
\end{split}
\end{equation}


\begin{figure*}
\centering
  \includegraphics[width=0.9\textwidth]{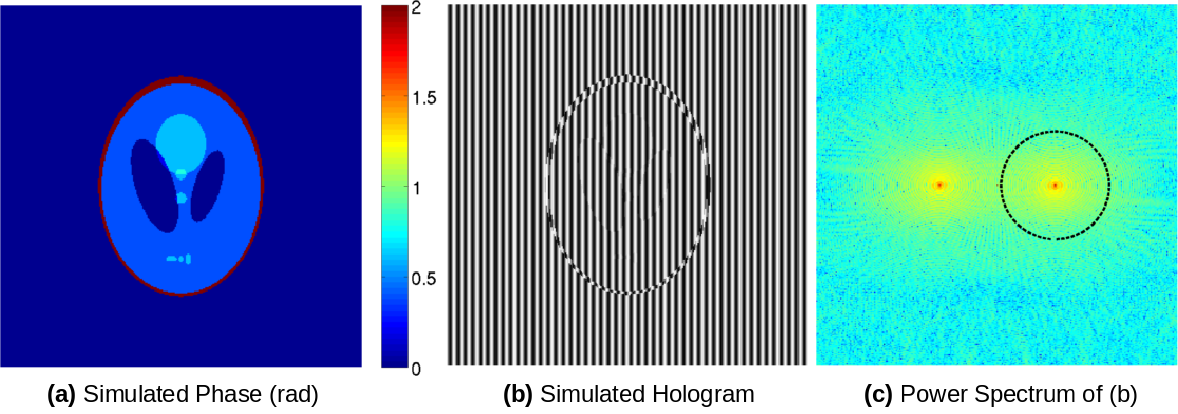}
\centering
  \caption{Reconstruction by Fourier filtering on simulated data. (a) Simulated phase image (rad). (b) Simulated hologram using (a). (c) Power spectrum of (b).}
  \label{Figure:Phantom}
\end{figure*}

\begin{figure*}
\centering
  \includegraphics[width=0.7\textwidth]{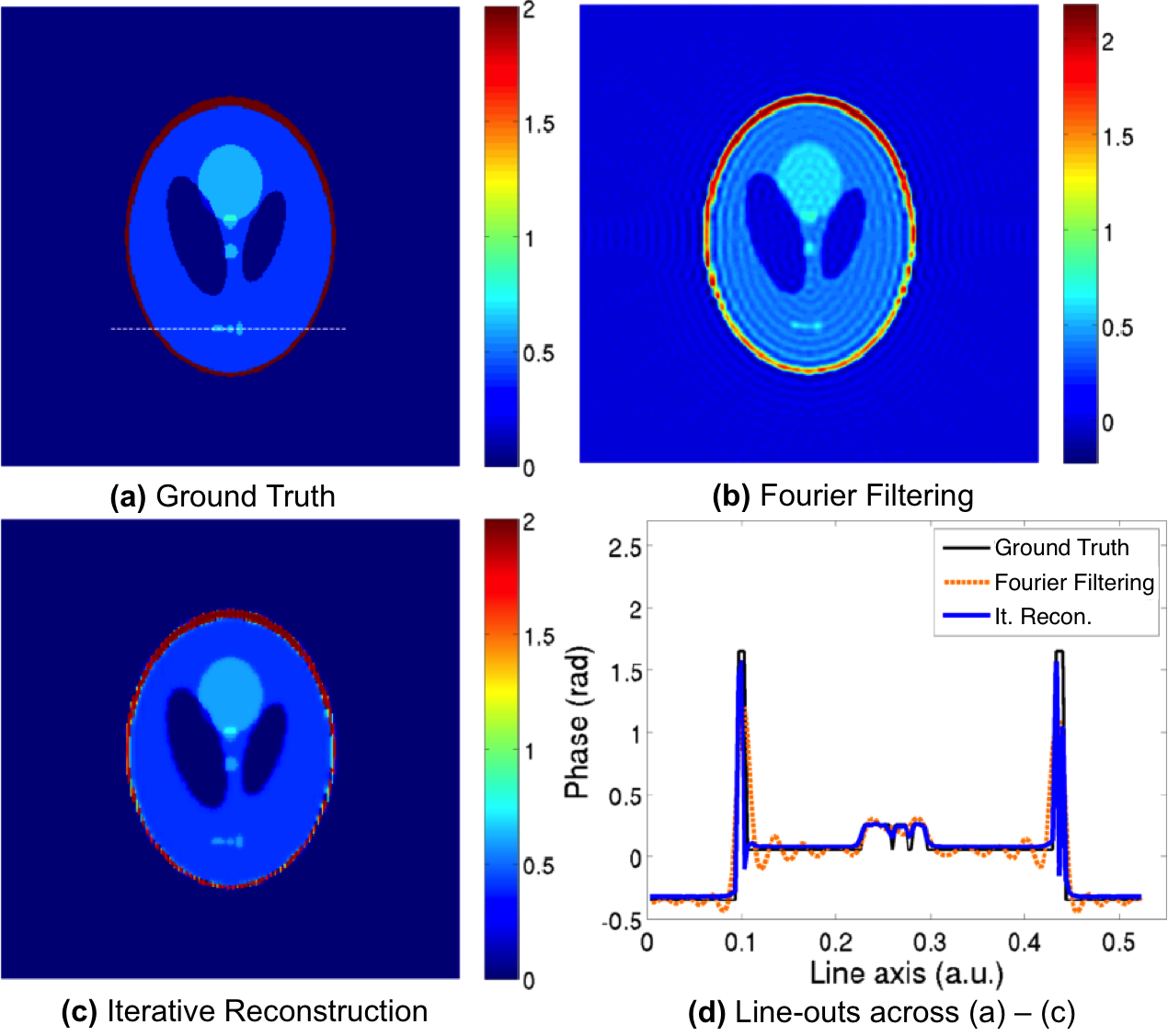}
\centering
  \caption{Phase reconstruction of the simulated phantom. (a) Ground truth phase (rad). (b) Reconstructed phase using Fourier filtering. (c) Reconstructed phase using iterative reconstruction, the proposed method. (d) Line-outs across the dashed white line in (a) with comparison to (b) and (c).}
  \label{Figure:Phantom_Reconstruction}
\end{figure*}

\begin{figure*}
\centering
  \includegraphics[width=0.7\textwidth]{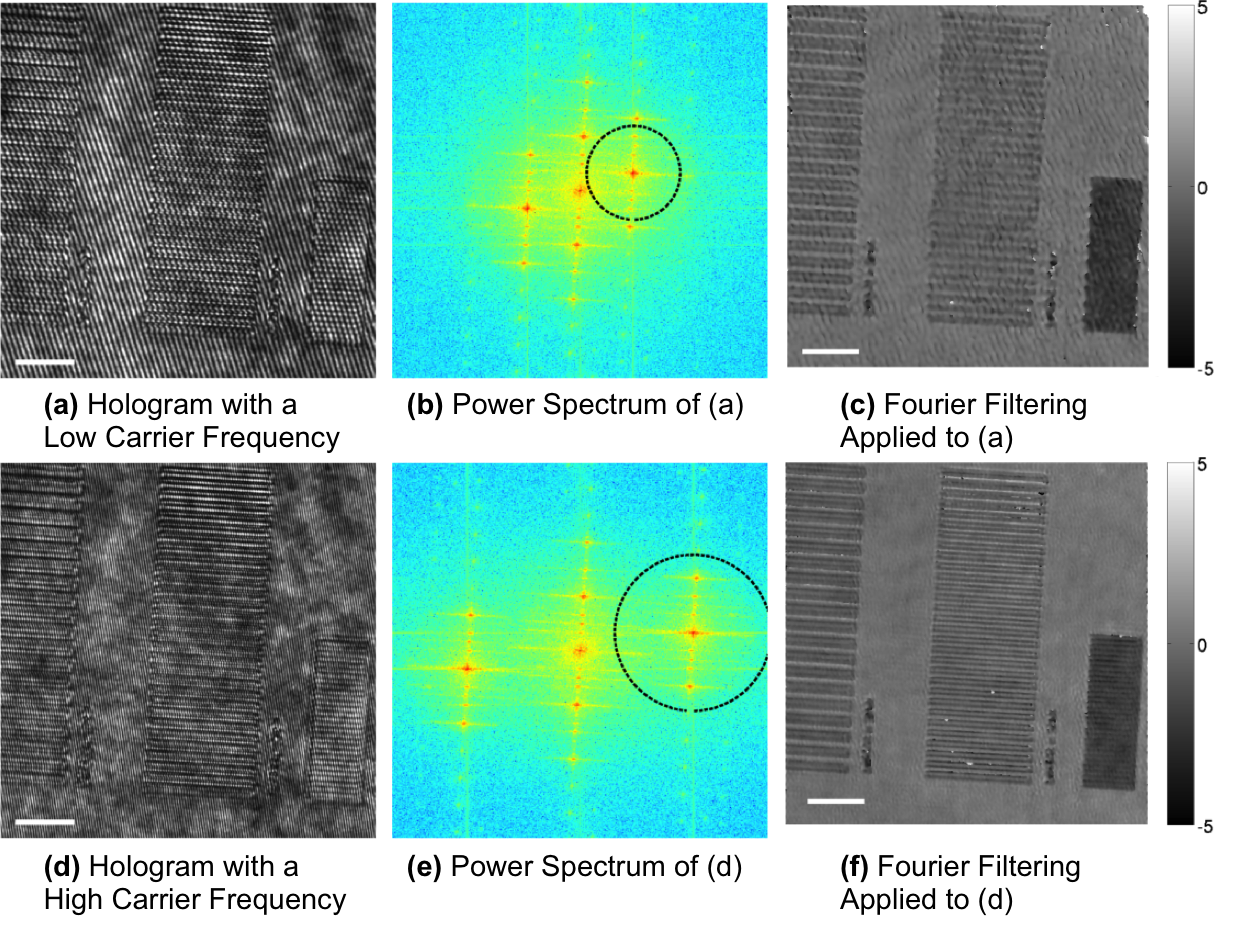}
\centering
  \caption{Reconstruction by Fourier filtering on experimental data. (a) Hologram with a low carrier frequency. (b) Power spectrum of (a). (c) Reconstructed phase using Fourier filtering on the low carrier frequency hologram. (d) Hologram with a high carrier frequency. (e) Power spectrum of (d). (f) Reconstructed phase using Fourier filtering on the high carrier frequency hologram. The scale bar lengths are $ 10 ~\mu m $.}
  \label{Figure:Hologram}
\end{figure*}

\begin{figure*}
\centering
  \includegraphics[width=0.9\textwidth]{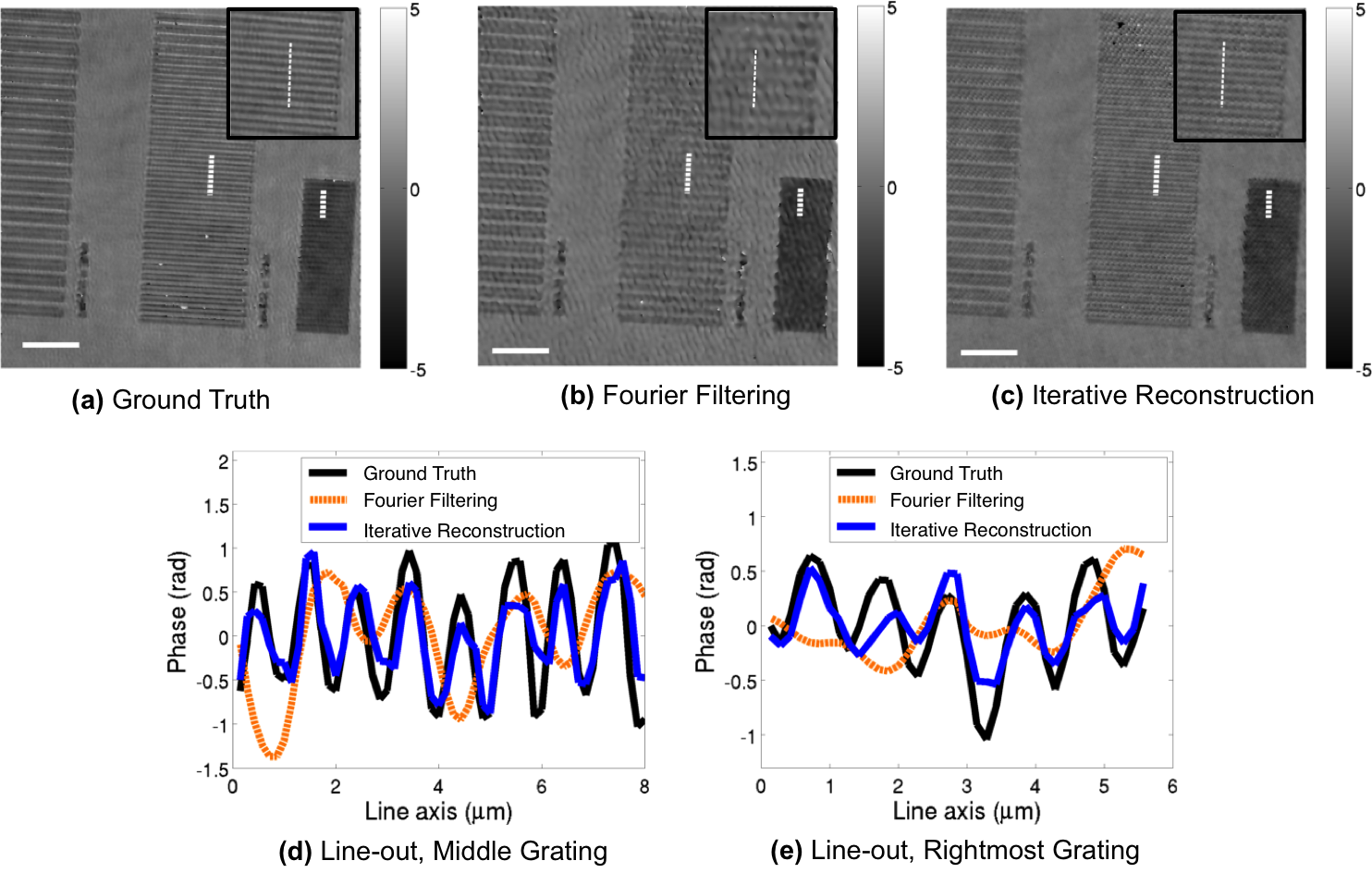}
\centering
  \caption{Phase reconstruction of a grating sample. (a) Ground truth phase (rad). (b) Reconstructed phase using Fourier filtering on the low carrier frequency hologram. (c) Reconstructed phase using iterative reconstruction, the proposed method, on the low carrier frequency hologram. (d), (e) Line-outs across the middle and rightmost gratings in (a)-(c). The scale bar lengths are $  10 ~\mu m $.}
  \label{Figure:Hologram_Reconstruction}
\end{figure*}

\section{Experiment}
\subsection{Simulation Results}

For simulation, we create a test object consisting of uniform amplitude $ A(\mathbf{x}) $, equal to 1 everywhere.  We set the phase $ \phi(\mathbf{x}) $ to have the form of the Shepp-Logan phantom,  shown in Fig. \ref{Figure:Phantom}(a).  The resulting object field, $ o(\mathbf{x}) = e^{i \phi(\mathbf{x})} $, may simulate a nearly transparent biological sample like a cell.  In an off-axis configuration, the reference beam is $ r(\mathbf{x}) = e^{i 2\pi \mathbf{f} \cdot \mathbf{x}} $.  We set the spatial frequency $ \mathbf{f} = (0.16, 0) $, with units of (pixel)$^{-1}$.  Figure \ref{Figure:Phantom}(b) shows the interference pattern $ I(\mathbf x) $ produced by $ o(\mathbf{x}) $ and $ r(\mathbf{x}) $. The Fourier filtering technique applies a filter to the Fourier transform of $ I(\mathbf x) $ to reconstruct the object. Figure \ref{Figure:Phantom}(c) highlights the filter, indicated by the black circle, superimposed on the power spectrum of $ I(\mathbf x) $.

Figure \ref{Figure:Phantom_Reconstruction} shows the phase reconstruction results, where we compare Fourier filtering and our iterative reconstruction approach as described in Table 1. The phantom in Fig. \ref{Figure:Phantom_Reconstruction}(a), the ground truth, has a high phase value of 2 rad at the outer edge, which contributes to the high frequency content in the power spectrum in Fig. \ref{Figure:Phantom}(c). Since the filter cuts off high frequencies, ringing occurs in the retrieved phase in Fig. \ref{Figure:Phantom_Reconstruction}(b). Our iterative reconstruction algorithm overcomes the limitations of filtering while reducing noise, as demonstrated in the reconstruction in Fig. \ref{Figure:Phantom_Reconstruction}(c). As a more quantitative comparison, we plot line-outs across the dashed white line in Fig. \ref{Figure:Phantom_Reconstruction}(a) and across the same points in Figs. \ref{Figure:Phantom_Reconstruction}(b) and \ref{Figure:Phantom_Reconstruction}(c).  The line-out in Fig. \ref{Figure:Phantom_Reconstruction}(d) displays the ringing in the phase computed by Fourier filtering, contrasting with the proposed method, which exhibits greater fidelity to the true phase and a better ability to resolve finer features.

\subsection{Experimental Results}

We test our proposed algorithm on experimentally measured data, which consists of holograms with low and high carrier frequencies as described in Table 2. Our sample consists of gratings patterned on PMMA film ($n$ = 1.49) with a height contrast of about 1.5 $ \mu $m. The camera measures an interference pattern, captured in Fig. \ref{Figure:Hologram}, with the reference beam at a slight angular offset from the object beam, under coherent illumination at 632.8 nm. Experimentally, we are able to vary the angular offset of the beams; a small offset leads to the hologram with a low carrier frequency in Fig. \ref{Figure:Hologram}(a), while a larger offset produces the hologram with a high carrier frequency in Fig. \ref{Figure:Hologram}(d). A higher carrier frequency enables larger sideband separation in the power spectra, evident in a comparison of Figs. \ref{Figure:Hologram}(b) and \ref{Figure:Hologram}(e), where the frequency content of the gratings appear aligned along the grating vectors. 

In the Fourier filtering technique, a filter isolates one of the sidebands, depicted as black circles in Figs. \ref{Figure:Hologram}(b) and \ref{Figure:Hologram}(e), for object reconstruction. The filter may not be able to capture the entire frequency content of the object, as in Fig. \ref{Figure:Hologram}(b), resulting in the blurrier reconstruction in Fig. \ref{Figure:Hologram}(c) compared to Fig. \ref{Figure:Hologram}(f). 

Figure \ref{Figure:Hologram_Reconstruction} compares reconstruction results. For ground truth, we measure a hologram with a high carrier frequency, from which we reconstruct the high resolution phase image in Fig. \ref{Figure:Hologram_Reconstruction}(a), using Fourier filtering. We measure a hologram with a low carrier frequency, our test data, with the goal of reconstructing a high quality image. Fourier filtering fails to capture high frequency details, as shown in Fig. \ref{Figure:Hologram}(b), resulting in the blurry reconstruction in Fig. \ref{Figure:Hologram_Reconstruction}(b). In contrast, our proposed method recovers the high frequency details, displayed in Fig. \ref{Figure:Hologram_Reconstruction}(c), that are lost with conventional Fourier filtering. Line-outs across the middle and rightmost gratings, plotted in Figs. \ref{Figure:Hologram_Reconstruction}(d) and \ref{Figure:Hologram_Reconstruction}(e), quantitatively demonstrate that our algorithm can reconstruct images in close agreement with the ground truth.

\section{Conclusion}
We have developed an image recovery approach to improve amplitude and phase reconstruction from single shot digital holograms, using iterative reconstruction with alternating updates. This approach allows the flexibility to apply different priors to amplitude and phase, improves phase reconstruction in image areas with low amplitudes, and does not require phase unwrapping for regularization. Phantom simulations and experimental measurements of a grating sample both demonstrate that the proposed method helps to reduce noise and resolve finer features. The improved image reconstruction from this technique will benefit the many applications of digital holography.


\begin{biography}
Dennis J. Lee received his Ph.D. in Electrical and Computer Engineering from Purdue University in West Lafayette, IN. He is a research scientist with Sandia National Laboratories in Albuquerque, NM. His research interests include computational imaging, signal and image processing, and optics.
\end{biography}

\end{document}